\documentclass{Interspeech}

\interspeechcameraready

\title{SDBench: A Comprehensive Benchmark Suite for Speaker Diarization}

\author[affiliation={1,2}, equalcontribution]{Berkin}{Durmus}
\author[affiliation={1}, equalcontribution]{Blaise}{Munyampirwa}
\author[affiliation={1}, equalcontribution]{Eduardo}{Pacheco}
\author[affiliation={1}, equalcontribution]{Atila}{Orhon}
\author[affiliation={1}, equalcontribution]{Andrey}{Leonov}

\affiliation{}{Argmax Inc}{USA}
\affiliation{}{University of California Los Angeles}{USA}

\email{eduardo@argmaxinc.com}

\usepackage{comment}

\begin{document}
\maketitle
\begin{abstract}
        \textcolor{black}{Even state-of-the-art speaker diarization systems exhibit high variance in error rates across different datasets, representing numerous use cases and domains. Furthermore, comparing across systems requires careful application of best practices such as dataset splits and metric definitions to allow for apples-to-apples comparison. We propose SDBench (Speaker Diarization Benchmark), an open-source benchmark suite that integrates 13 diverse datasets with built-in tooling for consistent and fine-grained analysis of speaker diarization performance for various on-device and server-side systems.
        \\
        SDBench\footnote{The code is available at \url{https://github.com/argmaxinc/SDBench}} enables reproducible evaluation and easy integration of new systems over time.
        To demonstrate the efficacy of SDBench, we built SpeakerKit, an inference efficiency-focused system built on top of Pyannote v3. SDBench enabled rapid execution of ablation studies that led to SpeakerKit being 9.6x faster than Pyannote v3 while achieving comparable error rates. We benchmark 6 state-of-the-art systems including Deepgram, AWS Transcribe, and Pyannote AI API, revealing important trade-offs between accuracy and speed.
        \\}
\end{abstract}

\begin{figure}[!ht]
  \centering
  \includegraphics[width=\linewidth]{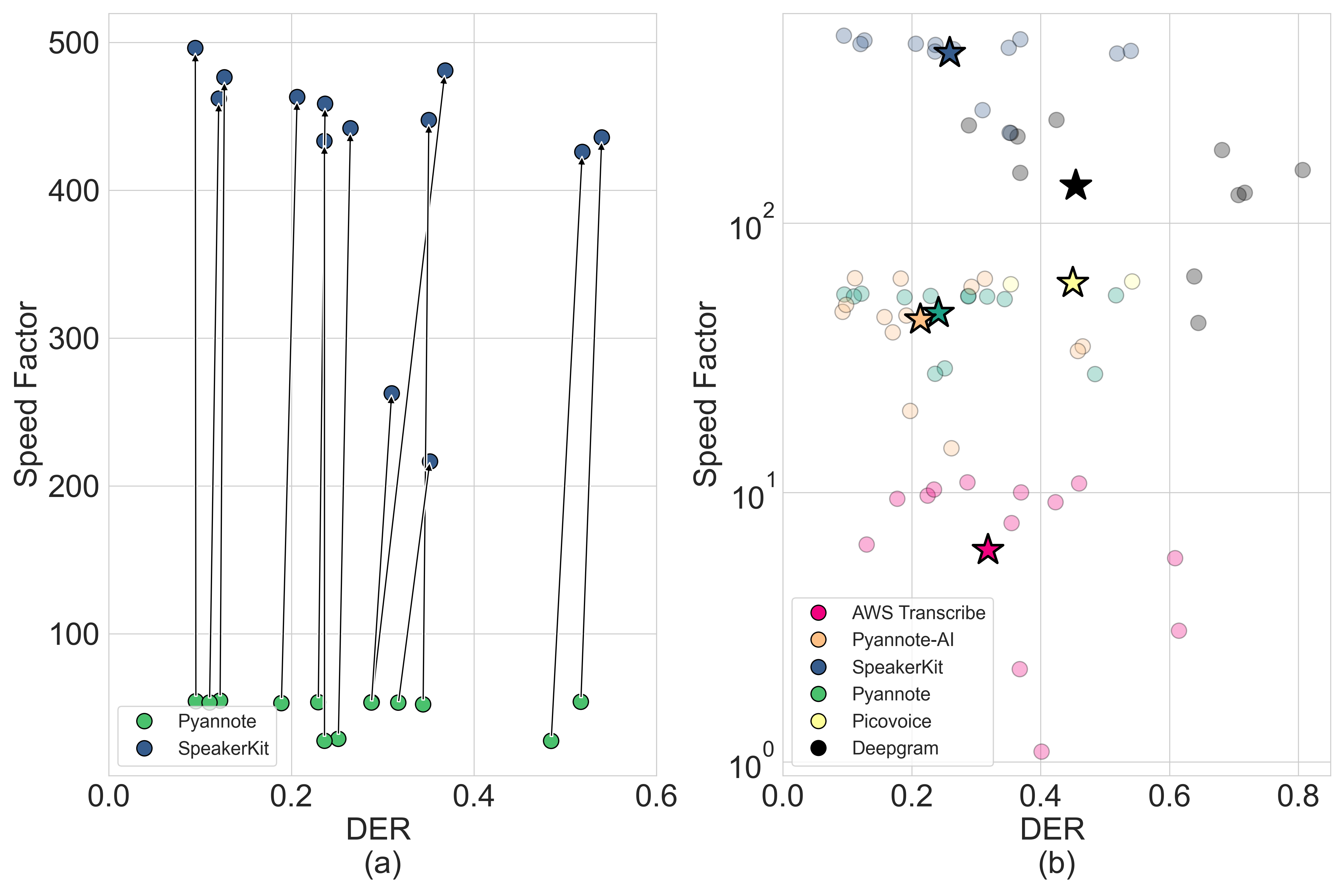}
  \caption{DER vs Speed Factor: (a) Pyannote v3.1 and SpeakerKit; SpeakerKit demonstrates a 9.6x speedup while achieving comparable DER, (b) 6 speaker diarization systems across 13 datasets. Circle markers represent per-dataset results whereas star markers represent cross-dataset aggregation.}
  \label{fig:der_vs_speed}
\end{figure}

\section{Introduction and Related Work}
Speaker diarization, the task of identifying "who spoke when" in audio data, is critical for applications such as meeting transcriptions, scribes, and voice assistants. Top-ranking systems on several recent challenges (\cite{diharddiarizationchallenge}, \cite{VoxCelebChallenge}, \cite{notsofarchallenge}) have adopted the following multi-stage architecture: (i) \textit{speaker segmentation} divides local audio windows into speaker-homogeneous segments, (ii) \textit{speaker embedding} extracts speaker-discriminative representations from each speaker segment, and finally (iii) \textit{clustering} groups segments by speaker identity based on speaker embeddings. Pyannote \cite{pyannote} is an open-source speaker diarization project that implements this multi-stage architecture. While this architecture lends itself to stage-targeted improvements, researchers would need advanced tooling for fine-grained error analysis and stage-wise evaluation to pursue such improvements.

State-of-the-art speaker diarization systems have recently reached commercially useful quality, prompting companies to productionize them through server-side APIs (\cite{awstranscribe}, \cite{pyannote-api}, \cite{gladia}, \cite{deepgram}) as well as on-device frameworks (\cite{pyannote}, \cite{picovoiceSDK}). In the presence of numerous open-source projects and proprietary products, consistent and comprehensive benchmarks are key to contextualizing each system's trade-offs and relative performance. However, many of these systems publish either ad-hoc benchmarks or none at all: \cite{picovoiceBenchmarks} evaluates systems on a non-standard split of a single dataset without noting the precise version of any system. Furthermore, they provide varying levels of speaker count-related information to some systems and not others and skip the overlapped speech segments. \cite{pyannoteGH} published benchmarks for various versions of Pyannote; however, reproducing these benchmarks and adding new systems remains a significant effort for others. The absence of an open-source benchmark to consistently evaluate and compare systems over time across various domains and use cases hinders the adoption of these systems in production.
\\
Furthermore, open-source academic projects have so far focused on accuracy-related metrics while practical aspects such as system latency remain underexplored \cite{pyannoteGHissue}. In contrast, proprietary systems such as Picovoice \cite{picovoiceSDK} improve efficiency for practical on-device deployment at the cost of accuracy. The absence of a high-accuracy and high-efficiency system hinders the adoption of on-device speaker diarization in production.
\\
Our main contributions can be summarized as follows:
\begin{itemize}
  \item SDBench, an open-source speaker diarization benchmark toolkit designed for fine-grained and consistent evaluation. Using SDBench, we publish benchmarks for 6 systems across 13 datasets and conduct ablation studies to justify various design decisions in Pyannote.
  \item SpeakerKit, a speaker diarization system built on top of Pyannote v3 that improves inference efficiency while retaining comparable error rates. SpeakerKit's development was guided by ablation experiments in SDBench, proving its efficacy in stage-targeted improvements.
\end{itemize}

While we benchmark several systems, this paper does not exhaust all available solutions, and we expect the community to integrate other state-of-the-art systems like NVIDIA NeMo \cite{nemo} and VBx \cite{vbx} as they adopt SDBench.

\section{Experiments}
\subsection{Datasets}
SDBench integrates 13 datasets spanning public and proprietary sources that represent multiple languages, audio domains, and speaker distributions: CALLHOME \cite{callhome}, DIHARD-III \cite{dihard3}, ICSI \cite{icsi}, Earnings-21 \cite{earnings21}, MSDWild \cite{msdwild}, AMI-IHM \cite{ami}, AMI-SDM \cite{ami}, VoxConverse \cite{voxconverse}, AliMetings \cite{alimeeting}, AISHELL-4 \cite{aishell4}, American-Life-Podcast \cite{americanlifepodcast}, AVA \cite{AVA}, and EGO4D \cite{ego4d}. Whisper large-v3 \cite{whisperkit} was used to predict the language spoken in each audio file to enable per-language evaluations. In this paper, datasets are marked as English, Chinese, or Multilingual for brevity.
\\
\\
Figure 2 presents key statistics of the test splits of each dataset: (a) total audio length, (b) overlap ratio, (c) speaker congestion, and (d) median speaker count. ICSI contains the longest total audio duration at ~70 hours, while CALLHOME \cite{callhome} and AMI \cite{ami} variants contain less than 10 hours each. AliMeetings \cite{alimeeting} shows the highest overlap ratio (0.19) among all datasets, indicating frequent speaker overlaps, while Earnings-21 \cite{earnings21} and American-Life-Podcast \cite{americanlifepodcast} show minimal overlap, reflecting their more structured nature. 
Many speaker diarization systems assume a maximum of 3 speakers per local audio window during the \textit{speaker segmentation} stage (\cite{pyannote}, \cite{wavlm}) and offload the burden of a higher number of speakers to later stages. However, \textit{speaker congestion}, measured as the percentage of windows that violate this maximum local speaker count assumption, is notably high in datasets like AliMeetings \cite{alimeeting} and ICSI \cite{icsi}, rendering these datasets predictably challenging. The median speaker count varies significantly, from 2 speakers in CALLHOME \cite{callhome} and DIHARD-III \cite{dihard3} to 19 in American-Life-Podcast \cite{americanlifepodcast}, enabling comprehensive evaluation across different speaker cohort size scenarios.
\\
\\
We published SDBench-compatible versions of these datasets on the Hugging Face Hub \cite{huggingface} 
\footnote{Hugging Face link will be available in the camera-ready version.} except for non-redistributable paid datasets (\cite{callhome} \cite{dihard3}). SDBench out-of-the-box includes scripts to reproduce the processing of all of the 13 original datasets.

\begin{figure}[t]
  \centering
  \includegraphics[width=\linewidth]{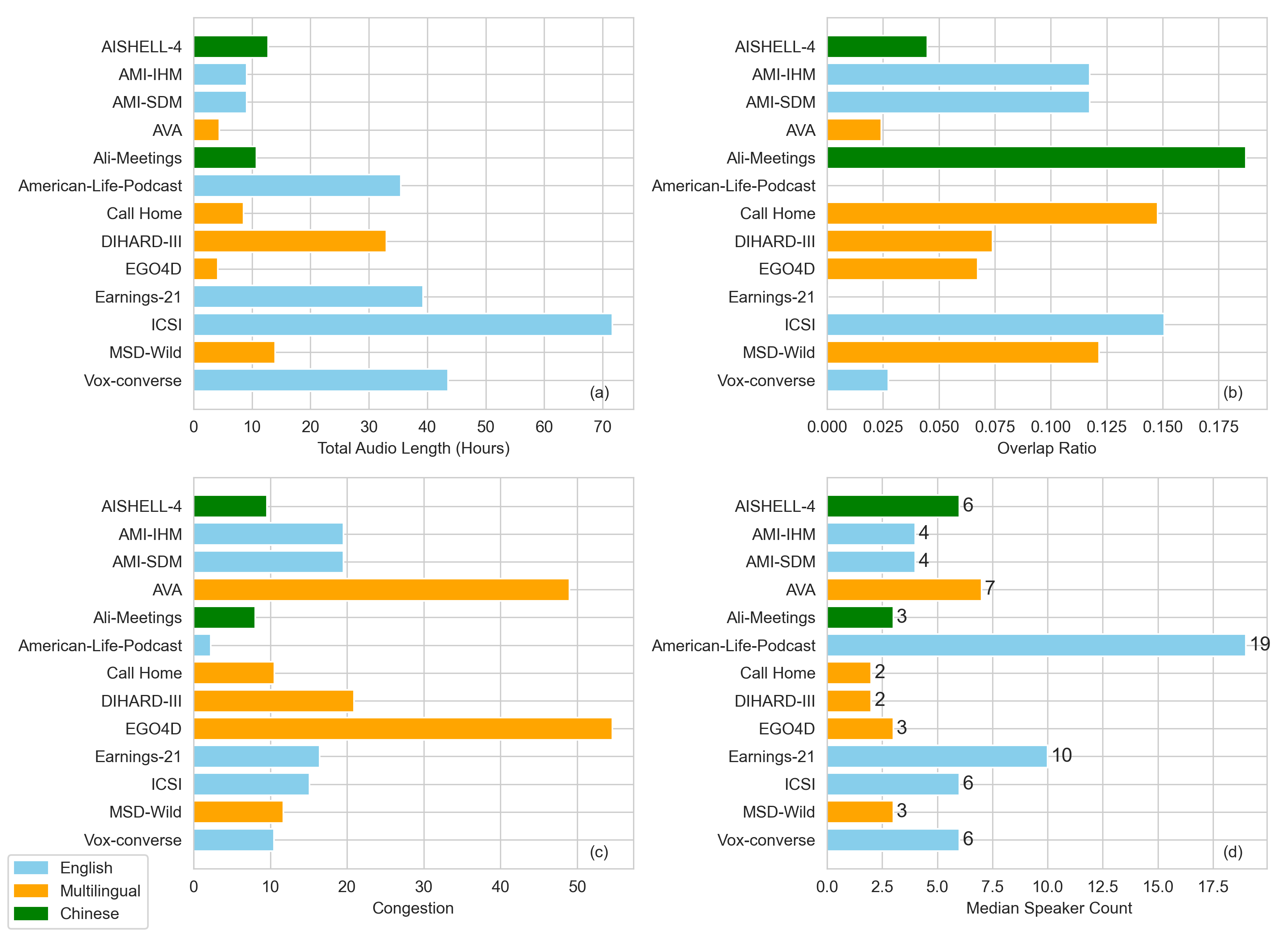}
  \caption{Key statistics for datasets used in this paper. (a) Total audio length in hours, (b) Overlap ratio (proportion of audio with more than 1 active speaker), (c) Speaker congestion (ratio of sliding windows with more than 3 active speakers), and (d) Median speaker count.}
  
  \label{fig:dataset_stats}
\end{figure}

\begin{figure}[th!]
  \centering
  \includegraphics[width=\linewidth]{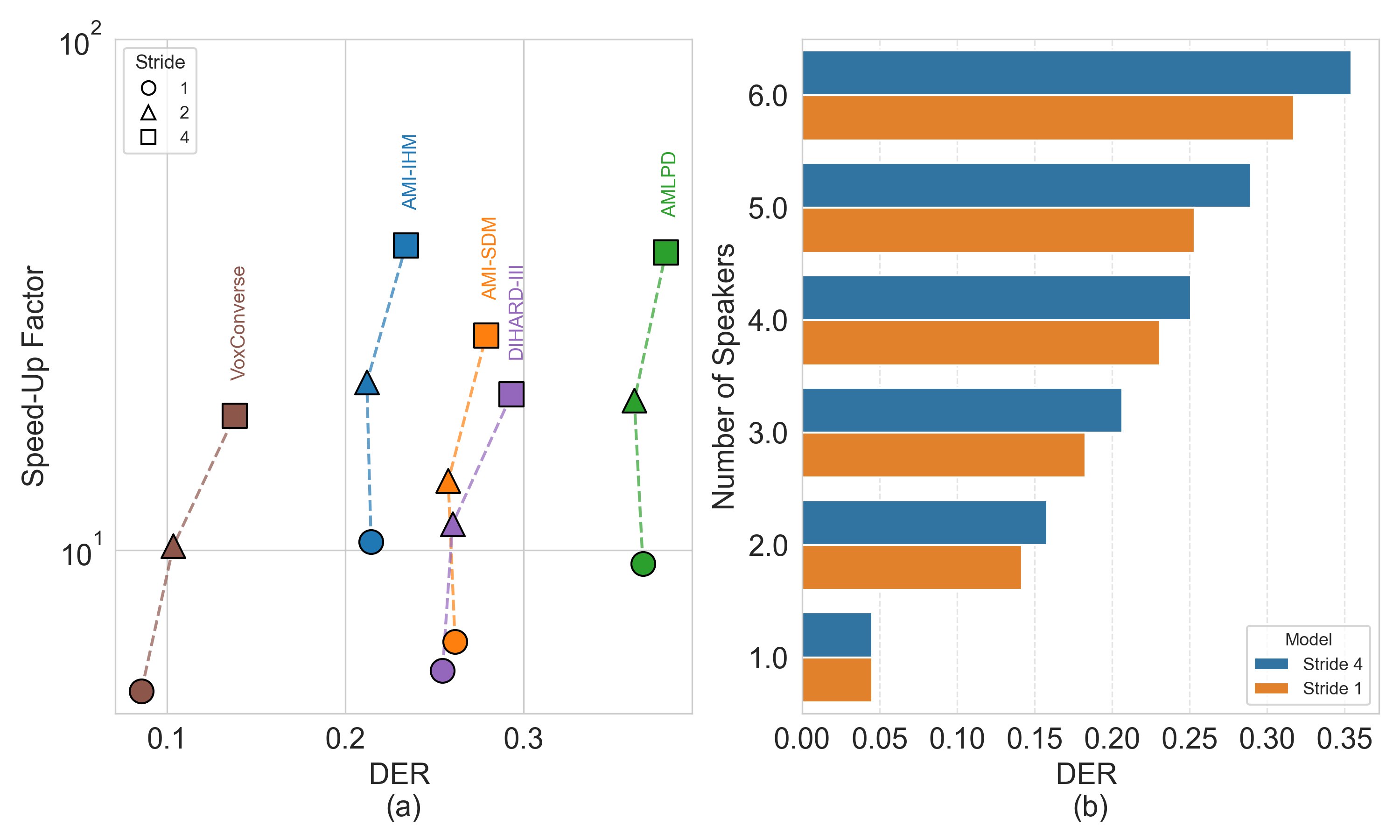}
  \caption{\textbf{Sliding window strategy}. (a) Relation between DER and Speed-up in pyannote units for strides 1, 2 and 4. (b) Shows the DER comparison between stride 1 and 4 across speaker counts. DER is impacted by less than 0.02 for up to 5 speakers and by less than 0.05 for unlimited number of speakers.}
  \label{fig:sliding_window}
\end{figure}

\subsection{Metrics}

\textbf{Diarization Error Rate (DER)}: This is the most common metric for evaluating speaker diarization systems. Unlike other benchmarking tools that use forgiveness collars depending on the dataset in DER calculation, we chose not to apply them across all datasets. However, adding forgiveness collars to SDBench for evaluation is straightforward if desired. Additionally, for computing DER, we used the global aggregation strategy defined in Pyannote \cite{pyannote}. DER is calculated as follows: 
\begin{equation}
\text{DER} = \frac{\text{False alarm} + \text{Missed detection} + \text{Speaker confusion}}{\text{Time Duration}}
\end{equation}
\\
\\
\textbf{Speed Factor}: is used throughout this paper to measure the speed of request completion for speaker diarization systems, calculated as:

\[
\text{Speed Factor} = \frac{ \text{Total Audio Duration}}{\text{Total Request Completion Time}}
\]
\\
In simpler terms, the Speed Factor indicates the number of seconds of audio the system can process per second of real-time.
\\
\\
Server-side system latency includes (i) the time taken to upload the audio file, (ii) polling  and download the results since this overhead is an inherent system feature and can not be circumvented in practice due to these systems' proprietary nature. Furthermore, if a system does not allow for speaker diarization requests independent of transcription requests (e.g. AWS Transcribe \cite{awstranscribe}, Deepgram \cite{deepgram}), the total request completion may have overhead from the required transcription workload.

\subsection{Ablation Studies}

\begin{figure}[th!]
  \centering
  \includegraphics[width=\linewidth]{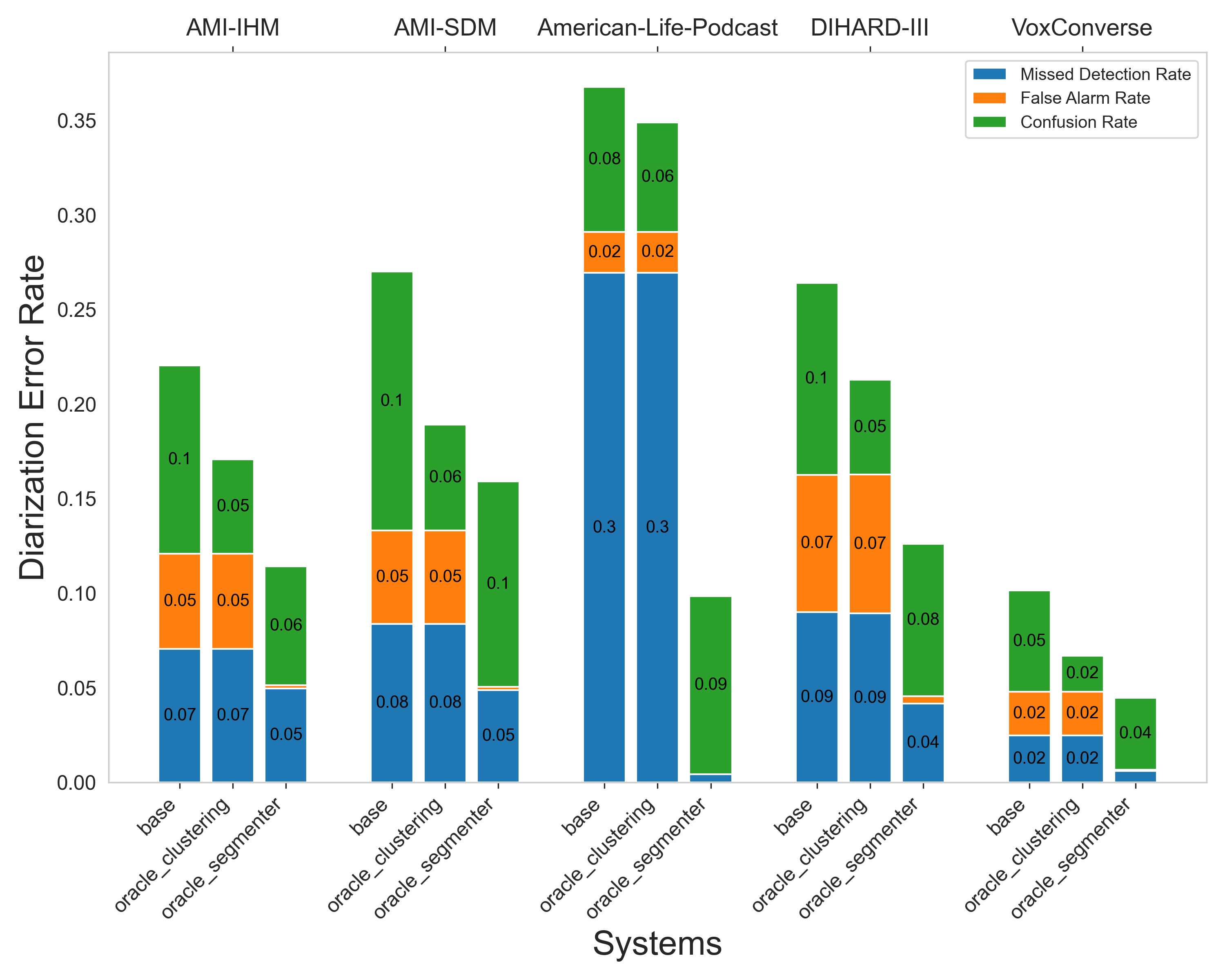}
  \caption{\textbf{Stage-wise evaluations.} Oracle Segmenter and Oracle Clusterer systems are compared to the standard Pyannote v3.1 system (Base) without any idealized components across 5 datasets. DER is broken across Missed Detection, False Alarm, and Confusion.}
  
  \label{fig:stagewise_eval}
\end{figure}

We pick Pyannote v3.1 as the reference system and conduct ablation studies to (i) demonstrate the breakdown of error rates across various stages and (ii) measure the impact of perturbing various design decisions. We use two short-form datasets — VoxConverse \cite{voxconverse} and DIHARD-III \cite{dihard3}—and three long-form datasets—American-Life-Podcast, AMI-SDM \cite{ami}, and AMI-IHM \cite{ami}. We use the validation splits to avoid overfitting on the test splits.
\\
\\
\textbf{Stage-wise evaluations.} We implement and evaluate the Oracle Segmenter system which simply feeds the ground truth speaker segments into the \textit{speaker embedding} stage. This idealized system is useful for measuring the minimum attainable error rates assuming perfect \textit{speaker segmentation} in each local window. Moreover, we implement and evaluate the Oracle Clusterer system which simply assigns the ground truth speaker identities to each predicted speaker segment using the minimum cost assignment regardless of the predicted speaker embeddings. This idealized system is useful for measuring the minimum attainable error rates assuming the best possible \textit{speaker identity} prediction based on predicted speaker segments. These idealized systems offer insights into where the next major error rate reductions may come from across various domains. However, they are not well-suited for performing isolated A/B tests for proposed implementation changes. Complex interactions between different stages may lead to end-to-end results contradicting isolated A/B tests. The results are shown in Figure~\ref{fig:stagewise_eval}.
\\
\begin{table}[th!]
  \caption{Speaker Embedder Ablations}
  \label{tab:word_styles}
  \centering
  \begin{tabular}{|p{3cm}|p{1cm}|p{2cm}|}
    \hline
    \textbf{Embedder} &
    \textbf{DER}    &  
    \textbf{Speed Factor}  
    \\
    \hline
    Pyannote (per window)&0.255        &   373   
    \\
    \hline
    Pyannote (per chunk) & 0.257         & 446
    \\
    \hline 
  \end{tabular}
\end{table}
\textbf{Sliding window strategy.} Pyannote v3.1 performs sliding window inference which introduces high temporal prediction redundancy, i.e. 10-second window size and 1-second stride. This redundancy increases the system's robustness to random error by redundant prediction averaging at the cost of considerable computational overhead. We benchmark the impact of reducing this redundancy on DER and Speed Factor by using longer sliding window strides, i.e. 1, 2 and 4 seconds. The results are shown in Figure~\ref{fig:sliding_window}.
\\
\\
\textbf{Speaker embedding strategy.} As an example, Pyannote v3.1 executes the \textit{speaker embedding} model on a total of 630 seconds of (highly overlapping) 10-second audio windows for a 30-second audio input. This 21x overhead can be attributed to running this model independently for (i) each of the 3 speakers regardless of whether they were active in that window or not, and (ii) each sliding window to cover the full audio length. 

We note that the speaker embedding model \cite{wang2023wespeaker} was designed to be speaker-agnostic except for the final layer where the per-speaker binary masks are used for temporal statistic aggregation to yield a fixed-size speaker-specific embedding. Armed with this observation, we design the \textit{per-chunk} strategy and embed the entire input audio chunk (instead of batching individual overlapping windows for each speaker) in a single forward pass while reproducing the same sliding window pattern per speaker in the final layer by using the speaker masks. This yields a fully non-redundant speaker embedding stage, e.g. embedding exactly 30 seconds of audio for a 30-second audio input, and we ablate this design decision by recomputing Speed Factor and DER. The results are shown in Table 1.

\subsection{Primary Benchmark}

\begin{figure}[th!]
  \centering
  \includegraphics[width=\linewidth]{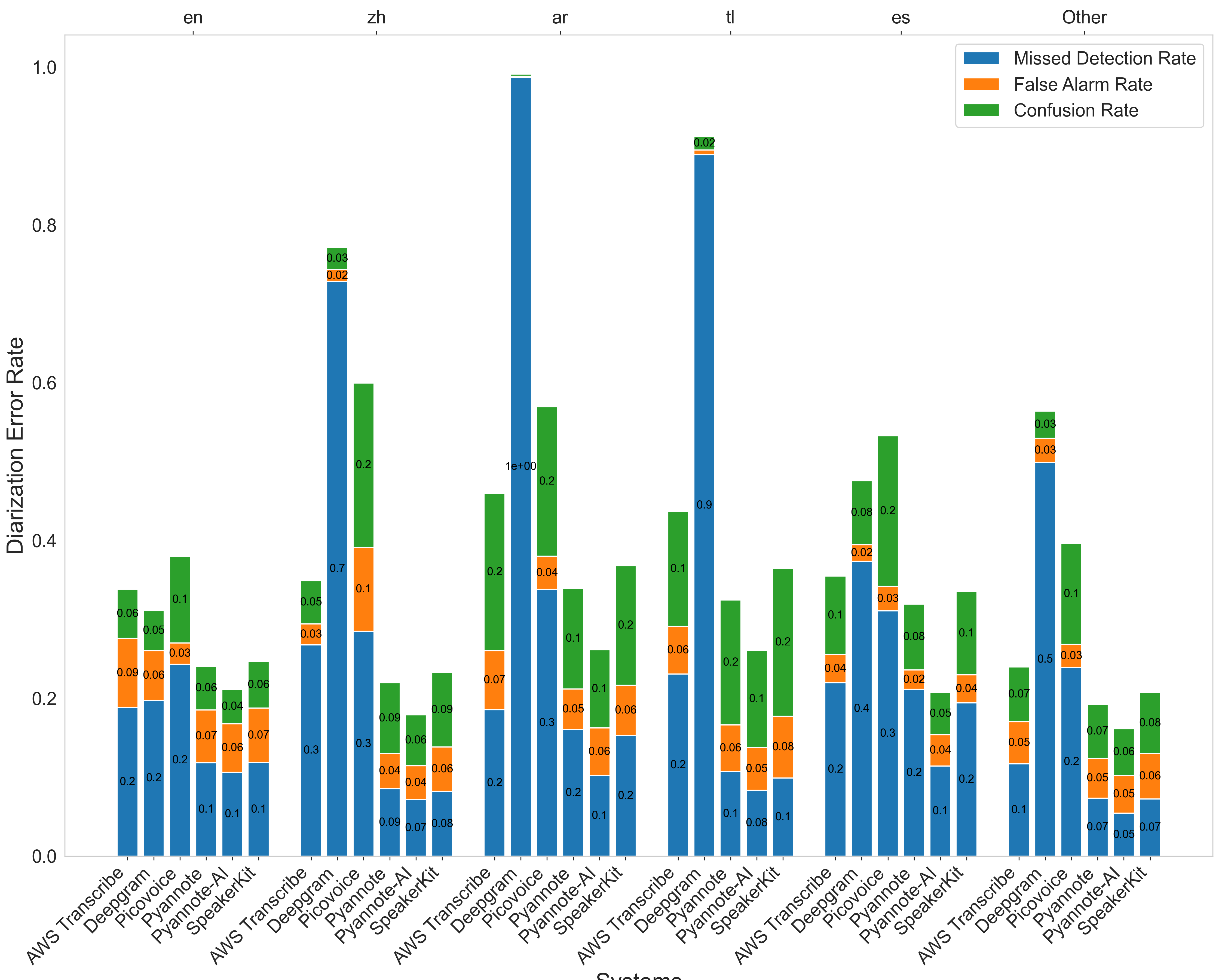}
  \caption{DER breakdown across multiple systems, categorized by language.}  

  \label{fig:der_by_language}
\end{figure}

\begin{figure}[th!]
  \centering
  \includegraphics[width=\linewidth]{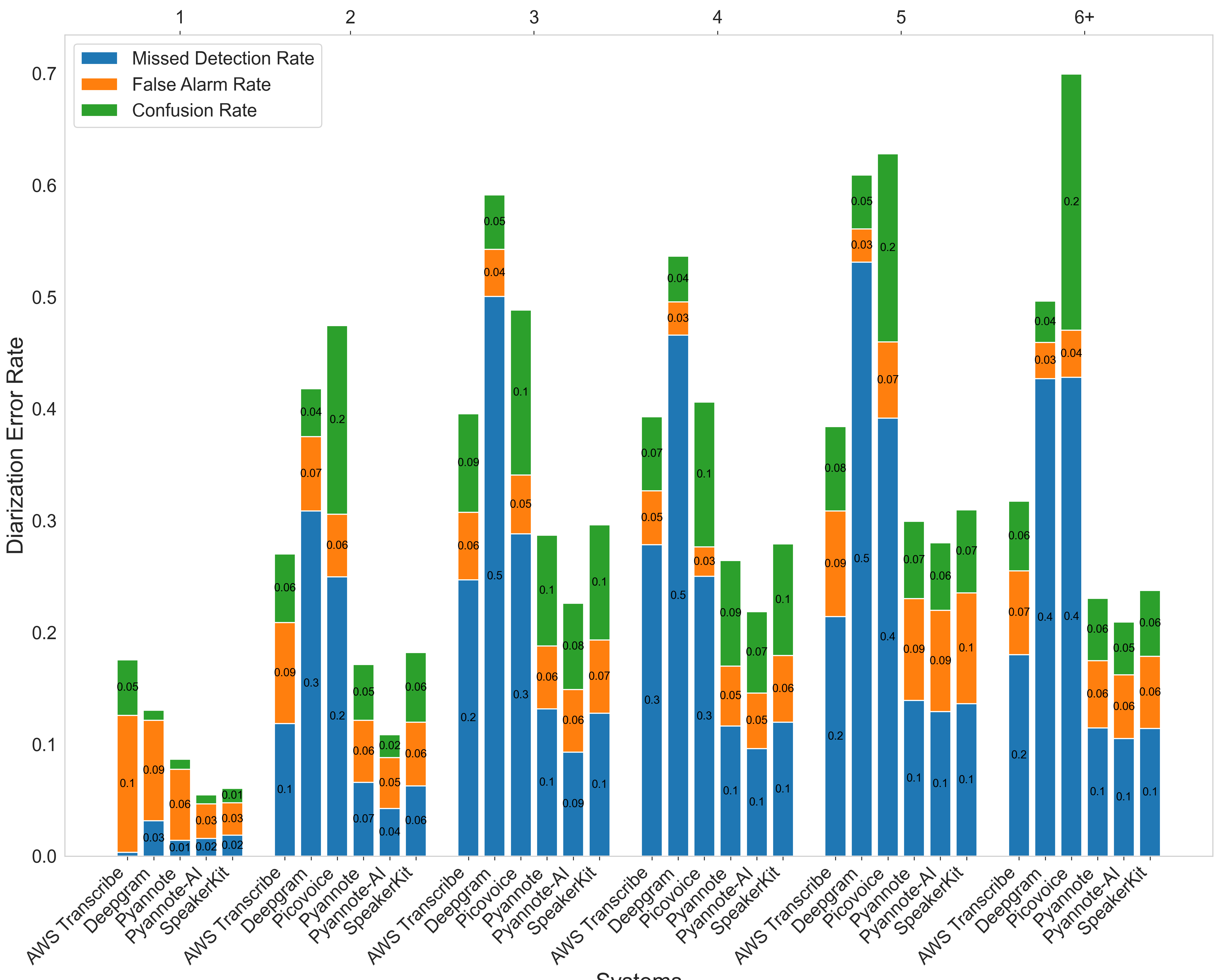}

  \caption{DER breakdown across multiple systems, categorized by number of speakers.} 
  \label{fig:der_by_speakers}
\end{figure}

We benchmark the following speaker diarization systems in the first version of SDBench:
\begin{itemize}
  \item \textbf{Pyannote v3.1} \cite{pyannote} via PyTorch MPS backend on M2 Ultra Mac Studio using float16 precision.
  \item \textbf{Picovoice Falcon v1.0.1 }\cite{picovoiceSDK} on M2 Ultra Mac Studio. The backend is not configurable.
  \item \textbf{AWS Transcribe} \cite{awstranscribe} v20250217 where versioning is based on the date benchmarks were conducted.\footnote{Access date-based versioning is necessary for server-side APIs due to the fact that underlying system changes are not guaranteed to be externally versioned.}
  \item \textbf{Deepgram} \cite{deepgram} v20250715 where versioning is based on the date benchmarks were conducted.\footnote{Due to API timeout errors, we were unable to benchmark Deepgram on the Earnings-21 and ICSI datasets.}
  \item \textbf{Pyannote AI API} \cite{pyannote-api} v20250217 where versioning is based on the date benchmarks were conducted.
  \item \textbf{SpeakerKit} v20250217, our proposed system built on top of Pyannote v3.1 running on M2 Ultra Mac Studio using float16 precision. Based on our ablation results, SpeakerKit is built on top of Pyannote v3.1 where the speaker embedding scheme is modified to per-chunk while maintaining stride 1 for the speaker segmenter.

\end{itemize}
3 of the 6 benchmarked systems leverage the multi-stage architecture described in Section 1 while the architecture of the other 3 systems is unknown. Figure~\ref{fig:der_vs_speed}b shows the evaluation results of these 6 systems on the test split of the 13 datasets described in Section 2.1. Please refer to the Appendix for a fine-grained breakdown of the results across speaker counts (Figure~\ref{fig:der_by_speakers}), languages (Figure~\ref{fig:der_by_language}), and other attributes.

\section{Results}
\textbf{Stage-wise evaluations.} Figure~\ref{fig:stagewise_eval} breaks down DER into 3 components: (i) Missed Detection, (ii) False Alarm, and (iii) Confusion. (i) and (ii) are attributed to the \textit{speaker segmentation} stage while (iii) is attributed to the \textit{clustering} stage. These results demonstrate that different error types are dominant across different datasets. However, the fine-grained breakdown in SDBenh enables researchers and practitioners to focus on stage-specific improvements for their system's intended use.
\\
\\
\textbf{Sliding window strategy.} Figure~\ref{fig:sliding_window} demonstrates that the sliding window stride significantly impacts Speed Factor while maintaining reasonable DER. Figure~\ref{fig:sliding_window} (a) shows dataset-specific impacts: DER differences range from 0.01 for AMI-IHM to 0.06 for VoxConverse at stride 4, while achieving speedups of ranging from 15.3x up to 38.3x compared to Pyannote. Figure~\ref{fig:sliding_window} (b) reveals that increasing stride from 1 to 4 impacts DER minimally: only 0.02 for up to 5 speakers and 0.05 for unlimited speakers. 
\\
\\
\textbf{Speaker embedding strategy.} Table 1 demonstrates that the \textit{chunk-level} strategy is an effective way to retain the same DER while achieving a 1.2x speedup in our SpeakerKit system. The speedup may be even higher in Pyannote v3.1.
\\
\\
\textbf{Primary benchmarks.} Figure~\ref{fig:der_vs_speed} demonstrates that Pyannote AI API attained the lowest DER. SpeakerKit attained the highest Speed Factor, despite running locally, while achieving comparable DER to Pyannote. Deepgram achieved the highest Speed Factor among server-side systems, though with higher DER compared to Pyannote AI API. AWS Transcribe attained the lowest Speed Factor which is likely related to inflexible system design, requiring speaker diarization requests to be added on top of transcription requests. Picovoice attained the highest DER \footnote{Due to usage limits in Picovoice Falcon free-tier, we were only able to benchmark this system on 2 out of 13 datasets: CALLHOME and AMI-IHM.} while achieving the second highest Speed Factor. 

\section{Discussion \& Conclusions}

We introduced SDBench, an open-source speaker diarization benchmark toolkit enabling fine-grained error analysis across various domains through stage-wise evaluations and 13 diverse datasets. To demonstrate its effectiveness, we developed SpeakerKit, an inference efficiency-optimized system built on Pyannote v3 that maintains low DER while achieving a 9.6x speedup. Our analysis revealed that sliding window redundancy benefits vary by domain, with controlled domains showing minimal DER impact (0.02) at stride 4 while achieving 30x speedups. The chunk-level speaker embedding approach achieved a 1.2x speedup without compromising DER performance. Leveraging SDBench, researchers and
practitioners can efficiently conduct ablation studies and continue improving their systems in similar spirit to how we built SpeakerKit.

\textbf{Scope and Limitations.} SDBench focuses on standalone diarization performance to enable clear analysis of fundamental challenges. While real-world applications often require joint optimization with tasks like ASR, such integration would introduce variables that could obscure core diarization performance. This focused approach allows researchers to optimize diarization independently before considering joint optimization. While this is not an exhaustive benchmark, we expect other state-of-the-art systems like NVIDIA NeMo \cite{nemo} and VBx \cite{vbx} to be integrated as the research community adopts SDBench.

\subsection{Future Work}
Future work includes deeper analysis of results to improve SpeakerKit's DER, tracking system evolution over time, exploring streaming diarization clustering strategies, and benchmarking integrated systems for joint speaker diarization and ASR. We also plan to expand SDBench with additional datasets and metrics.

\section{Appendix}
 
\bibliographystyle{IEEEtran}
\bibliography{mybib}

\end{document}